\documentclass[aps,prd,preprint,groupedaddress,superscriptaddress]{revtex4-1}
\usepackage[utf8]{inputenc}
\usepackage{amsmath}
\usepackage{amssymb}
\usepackage{epsfig}
\usepackage{soul}
\usepackage{graphicx}
\usepackage{color}
\usepackage[colorlinks=true
,urlcolor=blue
,anchorcolor=blue
,citecolor=blue
,filecolor=blue
,linkcolor=blue
,menucolor=blue
,pagecolor=blue
,linktocpage=true
,pdfproducer=medialab
,pdfa=true
]{hyperref}
\begin{document}


\title{Small field models of inflation \\ that predict a tensor-to-scalar ratio $r=0.03$}


\author{Ira Wolfson}
\email[]{irawolf@post.bgu.ac.il}
\affiliation{Department of physics, Ben-Gurion University of the Negev, 8410500 Beer-Sheva, Israel}

\author{Ram Brustein}
\email[]{ramyb@bgu.ac.il}
\affiliation{Department of physics, Ben-Gurion University of the Negev, 8410500 Beer-Sheva, Israel}
\affiliation{Theoretical Physics Department, CERN, 1211 Geneva 23, Switzerland}



\date{\today}

\begin{abstract}
Future observations of the cosmic microwave background (CMB) polarization are expected to set an improved upper bound on the tensor-to-scalar ratio of $r\lesssim 0.03$. Recently, we showed that small field models of inflation can produce a significant  primordial gravitational wave signal. We constructed viable small field models that predict a value of $r$ as high as $0.01$. Models that predict higher values of $r$ are more tightly constrained and lead to larger field excursions.  This leads to an increase in tuning of the potential parameters and requires higher levels of error control in the numerical analysis. Here, we present viable small field models which predict $r=0.03$. We further find the most likely candidate among these models which fit the most recent Planck data while predicting  $r= 0.03$. We thus demonstrate that this class of small field models is an alternative to the class of large field models.  The BICEP3 experiment and the Euclid and SPHEREx missions are expected to provide experimental evidence to support or refute our predictions.
\end{abstract}
\newpage


\maketitle

\section{Introduction}

Improved measurements of the B-mode polarization of the cosmic microwave background (CMB)  \cite{Seljak:1996ti,Seljak:1996gy} are expected to be more sensitive to the
tensor-to-scalar ratio $r$. This ratio provides a measure  of the amplitude of the primordial gravitational waves (GW),  which in turn is a telltale of inflation \cite{Lyth:1996im}.  The final Planck data release and analysis \cite{Aghanim:2018eyx} currently suggest an upper bound of $r < 0.07$. The BICEP experiment \cite{Ade:2014xna,Ade:2014gua,Ade:2018gkx,Barkats:2013jfa} took data over the last few years \cite{Wu:2016hul} which is expected to yield an upper bound of $r\leq 0.03$. A discovery of a value as high as $r \sim 0.03$  could indicate a very high energy scale $\sim 10^{16} GeV$ (see, for example, \cite{{Lyth:1998xn}}).

We continue our investigations \cite{Wolfson:2016vyx,Wolfson:2018lel} of a class of inflationary models that were proposed by Ben-Dayan and Brustein  \cite{BenDayan:2009kv} and were followed by  \cite{Hotchkiss:2011gz,Antusch:2014cpa,Garcia-Bellido:2014wfa}. This class of models is compatible with several fundamental physics considerations. Recently, interest in this class of models was revived by the discussion about the ``swampland conjecture", \cite{Lehners:2018vgi,Garg:2018reu,Kehagias:2018uem,Ben-Dayan:2018mhe} which suggests that small field models are favoured by various string-theoretical considerations (see \cite{Palti:2019pca} for a recent review).

In addition, for this class of inflationary models, high values of $r$ result in a scale dependence of the scalar power spectrum. Future experiments such as Euclid \cite{Amendola:2012ys}, and SPHEREx \cite{Dore:2014cca} aim to probe the running of the scalar spectral index $\alpha$ at the level of $10\%$ relative error. This is a major improvement  in comparison to the Planck bounds on $\alpha$ which are currently at the level of $\sim 75\%$ relative error. Such future measurements could provide additional constraints on our models.

\begin{figure}[!h]
\includegraphics[width=1\textwidth]{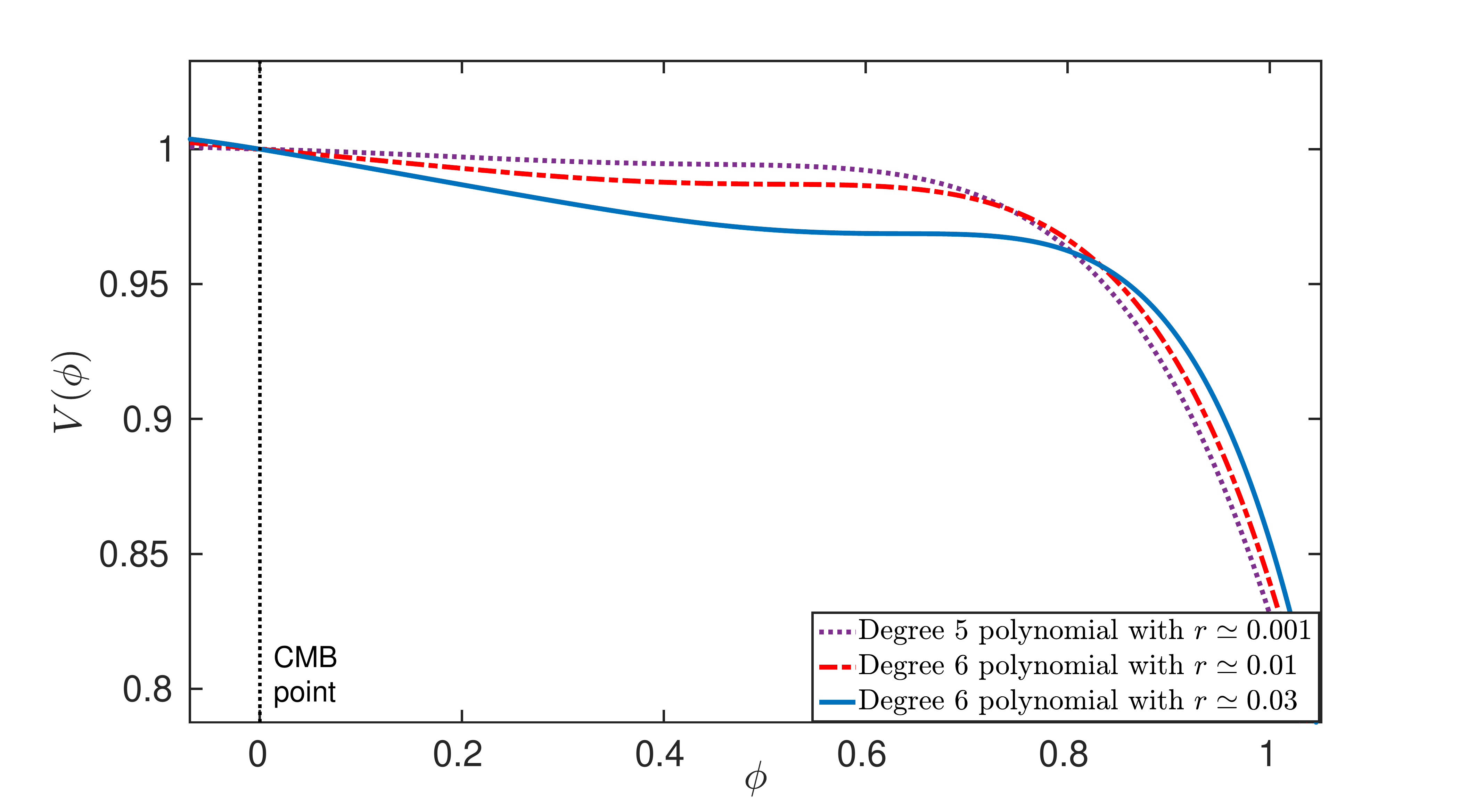}
\caption{Polynomial potentials. The blue (solid) line depicts a potential of a model that predicts $r\simeq 0.03$, while the red (dash-dotted) line depicts a potential that predicts $r\simeq 0.01$. The purple (dotted) line depicts a degree five polynomial potential of a model that predicts $r\simeq0.001$. All models are variants of the hilltop model, with a  flatter region in which most e-folds are generated.
\label{potentials}}
\end{figure}

\section{The models}

The small field models that we study are single-field models. The action of such models is given by
\begin{align}
	\mathcal{S}=\int d^4 x \sqrt{-g}\left[\frac{R}{2}-\frac{1}{2}\partial^{\mu}\phi\partial_{\mu}\phi -V(\phi)\right].
\end{align}
The metric is of the FRW form and the potential given by
\begin{align}
	V(\phi)=V_0\left[1+\sum_{p=1}^{6}a_p\phi^p\right].
\end{align}
Previously, in \cite{BenDayan:2009kv,Wolfson:2016vyx} this class of models was discussed from a phenomenological and theoretical points of view. In \cite{Wolfson:2016vyx}, the technical details of model building and simulation methods were discussed, while in \cite{Wolfson:2018lel}, the analysis and the extraction of the most probable model were discussed. Additionally, in \cite{Wolfson:2018lel}, the most likely model which yields $r=0.01$ was identified.

\begin{figure}[!h]
\includegraphics[width=1\textwidth]{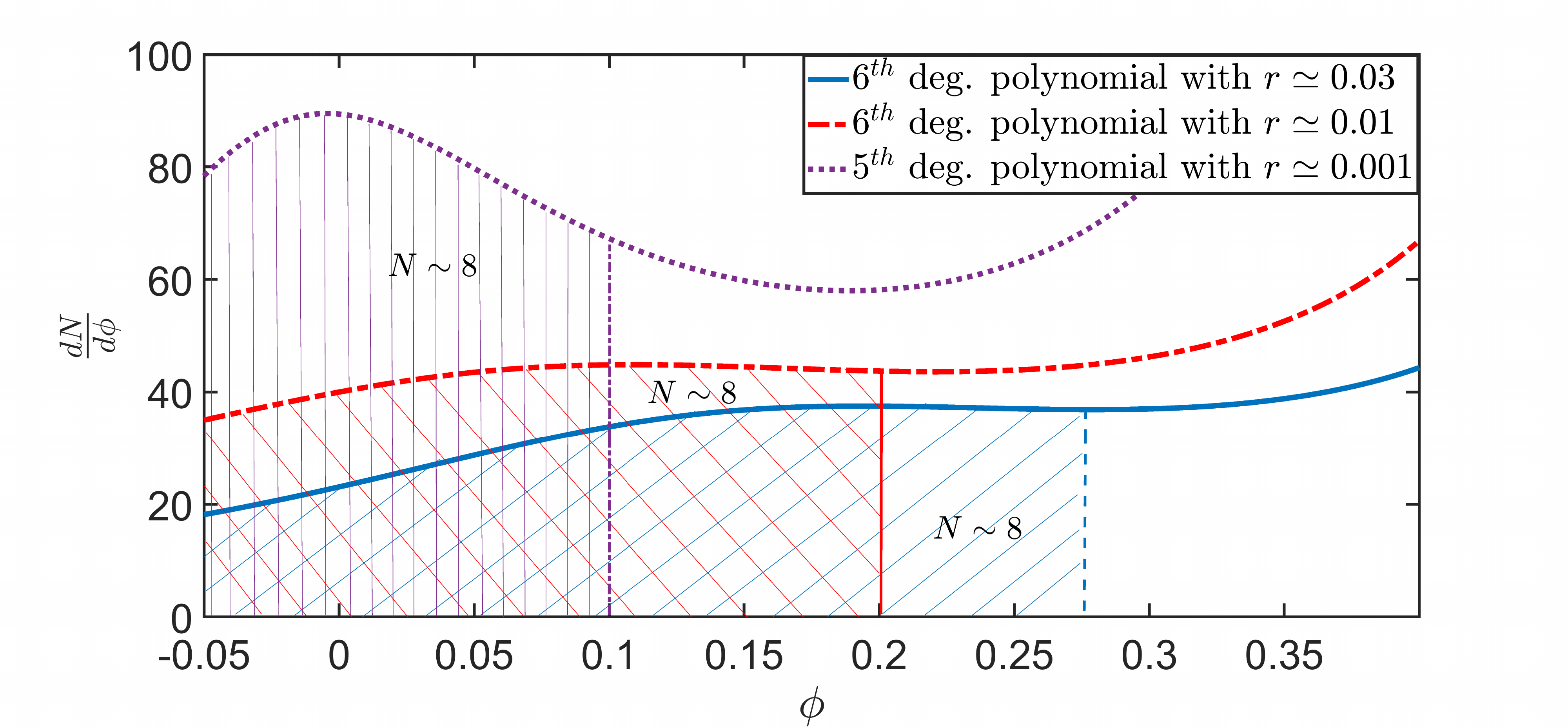}
\caption{Field excursions in (reduced) Planck units. Different predicted values of $r$ require different field excursions to generate the $\sim 8$ e-folds probed by the CMB. The model predicting $r\simeq 0.03$ (blue line) requires an excursion of $\left(\Delta\phi\right)_{CMB}\simeq 0.28$ to generate the same amount of e-folds which the model predicting $r\simeq 0.01$ model (red dash-dot) generates in $\left(\Delta\phi\right)_{CMB}\simeq 0.2$. This means more tuning is required for models that predict $r\simeq 0.03$ . The model predicting $r\simeq0.001$  (purple dots) requires only $\left(\Delta\phi\right)_{CMB}\simeq 0.1$ to generate the CMB window. \label{dNdphi}}
\end{figure}

\section{Methods}

We analyse the most recently available observational data \cite{Aghanim:2018eyx} by using CosmoMC \cite{Lewis:2002ah}, extracting the likelihood curves of the scalar index, its running and the running of running, $n_s,\alpha,\beta$ respectively. We then simulate a large number of inflationary models with polynomial potentials and calculate the primordial power spectrum (PPS) observables $n_s$, $\alpha$, $\beta$, that they predict. Each simulated potential  is assigned a likelihood by the combined likelihood of the observables that it yields, as discussed in detail in \cite{Wolfson:2018lel}. We restrict the models that we consider to those predicting a power spectrum that can be fitted well by a third degree polynomial. This corresponds to the scalar index $n_s$, the index running $\alpha$, and the running of running $\beta$. We do that by fitting the PPS and evaluating the fitting error
\begin{align}
	\Delta^2=\frac{1}{n}\sum_{k=1}^n\left[\log(P_S(k))-f(k)\right]^2,
\end{align}
where $f(k)$ is the fitting curve. The threshold for considering a specific model in our analysis is $\Delta^2<10^{-6}$.

An additional complication arises due to the higher amount of tuning that is required for these models. The coefficient $a_1$ is fixed as $a_1=-\sqrt{r/8}$ \cite{BenDayan:2009kv}, so when the value of $r$ is higher, then $dV/d\phi$ at the CMB point has a larger magnitude. This has the effect of decreasing the number of e-folds generated per field excursion interval. If $r$ is increased to $Cr$, then $\frac{dN}{d\phi}$ is decreased by a factor  $\sim \frac{1}{\sqrt{C}}$
close to the CMB point. Since the first 8 or so e-folds of inflation are fairly constrained by observations, the amount of freedom in constructing the potential is reduced. It follows that either greater tuning is required to construct valid potentials, or one should consider a higher degree polynomial as suggested in \cite{Hotchkiss:2011gz}. Ultimately the choice is a matter of practical convenience. We opted for using  sixth degree polynomials.

We employ two methods of retrieving the most likely polynomial potential. First,  we extract by marginalisation the most likely coefficients $\{a_p\}$. The other method amounts to performing a multinomial fitting of the coefficients as a function of the observables and then inserting the most likely observables to recover the corresponding coefficients. This method is explained in detail in \cite{Wolfson:2018lel}.

The `most likely model', is the model with a potential that generates the most likely CMB observables  $n_S\simeq 0.9687,\alpha\simeq 0.008,\beta\simeq 0.02$, as produced by the MCMC analysis of the most recent data available to date.
\begin{figure}[!h]
\includegraphics[width=0.8\textwidth]{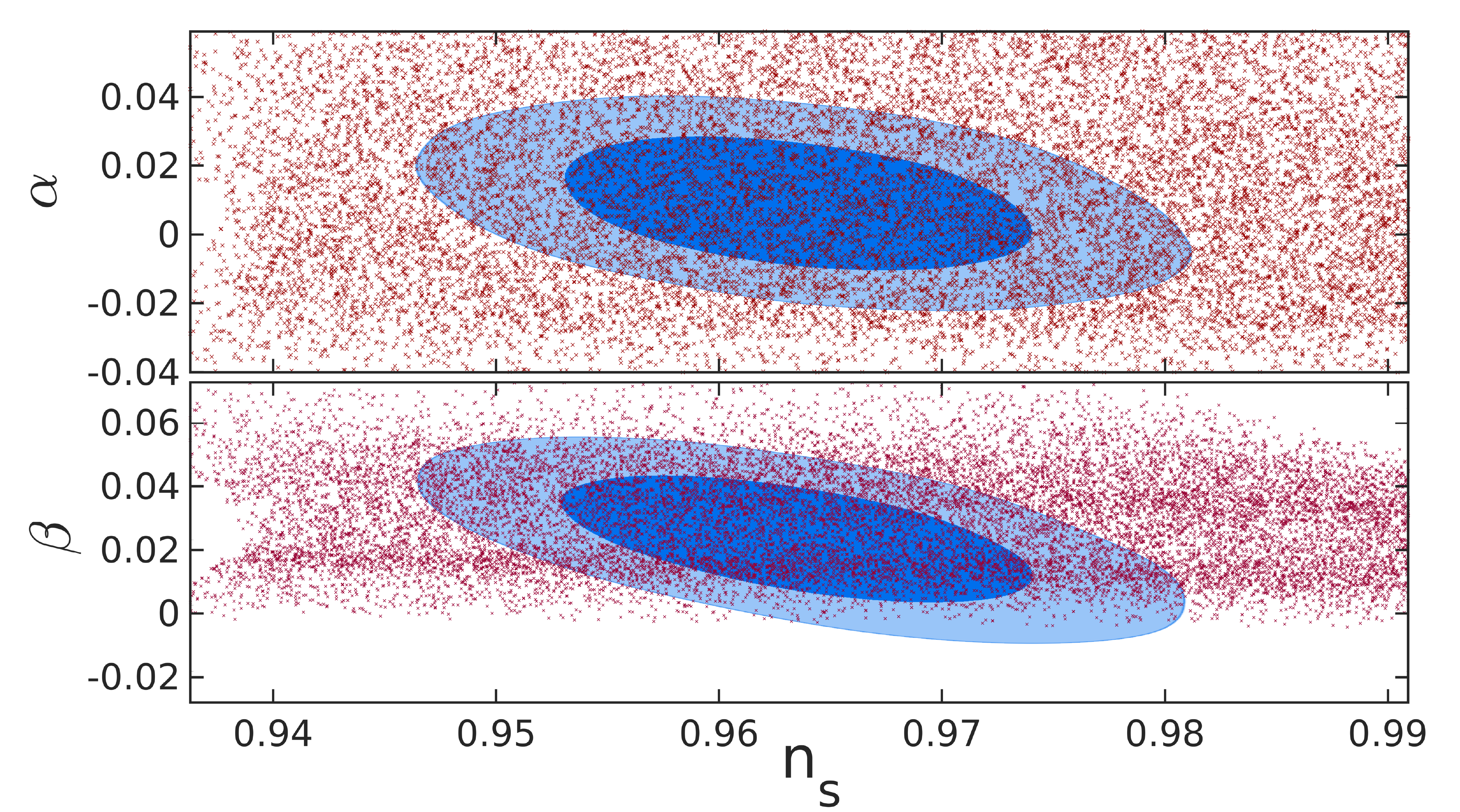}
\caption{Covering the observable phase space with small field models that predict $r\simeq 0.03$. The roughly uniform cover of the $95\%$ CL areas ensures an accurate likelihood transfer from MCMC results to models.\label{Main_result}}
\end{figure}

\section{Results}
\begin{figure}[!h]

\includegraphics[width=1\textwidth]{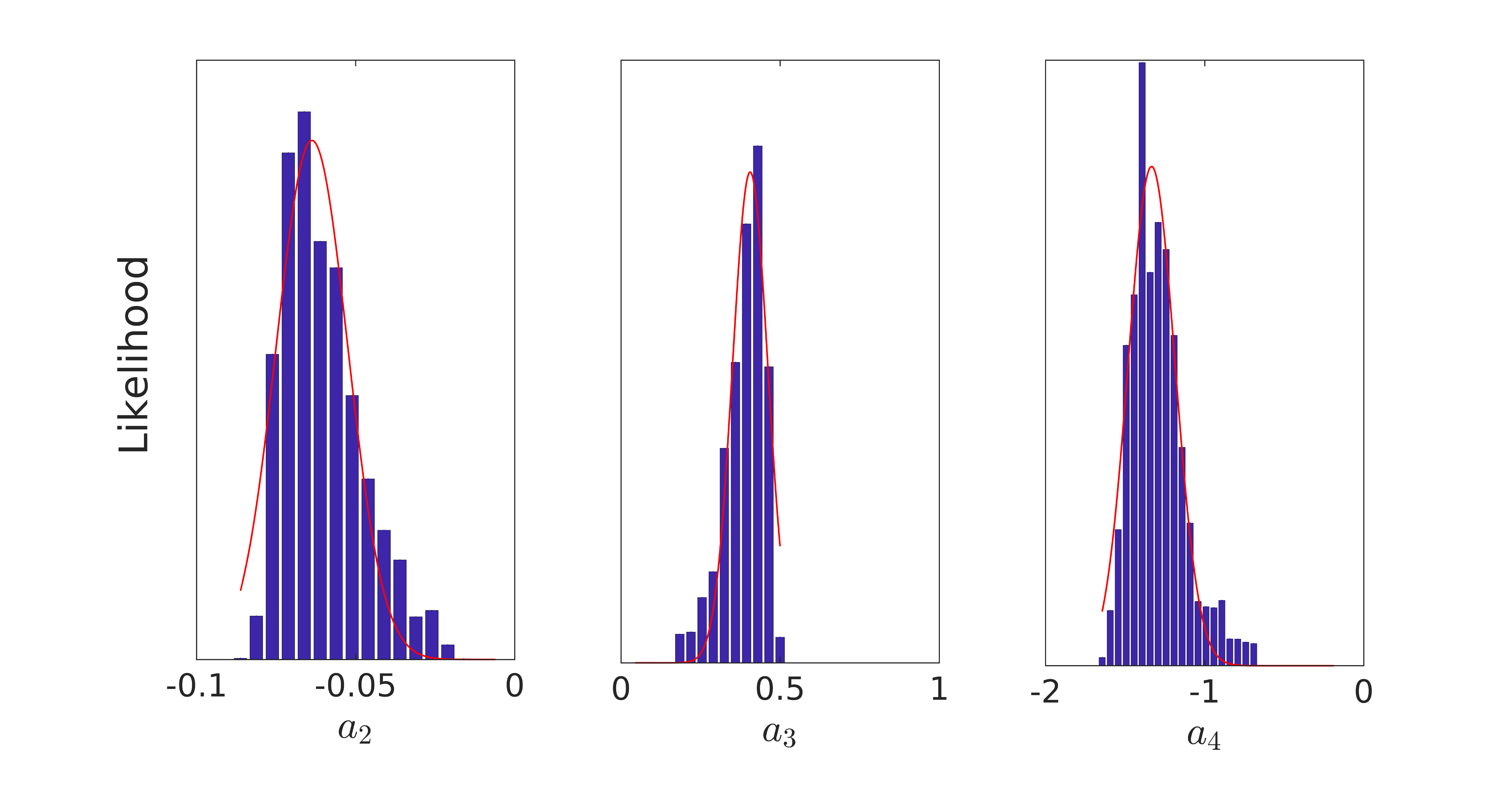}
\caption{Likelihood analysis for model coefficients $a_2,a_3,a_4$, using Gaussian fits to recover mean values. The skewness of the results reflect properties of the model, not of the underlying MCMC analysis. The width of the Gaussian fits are given by $(0.016,0.08,0.21)$ and are an indication of the tuning level required for these models.\label{Likelihood_results}}
\end{figure}

We produced many models that predict $r\simeq0.03$  and, additionally, predict PPS observables within the likely values. A roughly uniform cover of both the $(n_s,\alpha)$ and $(n_s,\beta)$ allowed values is shown in Fig.~\ref{Main_result}. This enables us to assign likelihood to each simulated model, as discussed in detail in \cite{Wolfson:2016vyx,Wolfson:2018lel}. Consequently, it is possible to perform a likelihood analysis of models.  The results of this likelihood analysis is an approximately Gaussian distribution of the free coefficients as is shown in Fig. \ref{Likelihood_results}. Since the peaks of Gaussian fit (red line in Fig.~\ref{Likelihood_results}) do not quite coincide with the peaks of the distribution and the distribution tails are not symmetric, we conclude that the distribution has a significant skewness.
The required tuning level is also evident from Fig.~\ref{Likelihood_results} and is given by $(\Delta a_2,\Delta a_3, \Delta a_4)=(0.016,0.08,0.21)$. Using Gaussian analysis and taking into account the skewness, we recover the most likely coefficients, which yield the following  degree six polynomial small field potential:
\begin{align}
	V=V_0\left[1-\sqrt{\frac{0.03}{8}}\phi -0.069\phi^2 + 0.431\phi^3 -1.413\phi^4 +2.455\phi^5 -1.487\phi^6\right]. \label{Most_likely_gauss}
\end{align}

Due to the skewness of the distribution, the values obtained by the Gaussian fit deviate by a significant amount from the most likely values of the observables. For instance, $n_s$ as determined by the potential in \eqref{Most_likely_gauss} is $\sim 0.98$ which is about $2\%$ away from the most likely value. For this reason we use this method of analysis to evaluate the required  tuning levels, whereas the most likely model is extracted by the multinomial fitting method.

Using the popular Stewart-Lyth (SL) theoretical values for $n_s$ and $\alpha$ \cite{Stewart:1993bc,Lyth:1998xn} as derived directly from the inflationary potential around the pivot scale, one finds  values that deviate by a large amount from the Planck values. The SL values correspond to a very blue power spectrum and large running,
\begin{align}
	\left.n_s\right|_{SL}\simeq 1.55 \\
	\nonumber \left.\alpha\right|_{SL}\simeq -0.216 .
\end{align}
This discrepancy, that was discussed in \cite{Wolfson:2016vyx}, is related to the magnitude of $r$ for our class of models. When $r$ is smaller than $\sim 10^{-4}$ in such models, the original model building procedure that relied on the SL values, which is outlined in \cite{BenDayan:2009kv}, is valid and produces approximately the correct values of the observables. However, when values of $r$ are larger, one cannot trust the analytic SL estimates.

\begin{figure}[!h]
\includegraphics[width=1\textwidth]{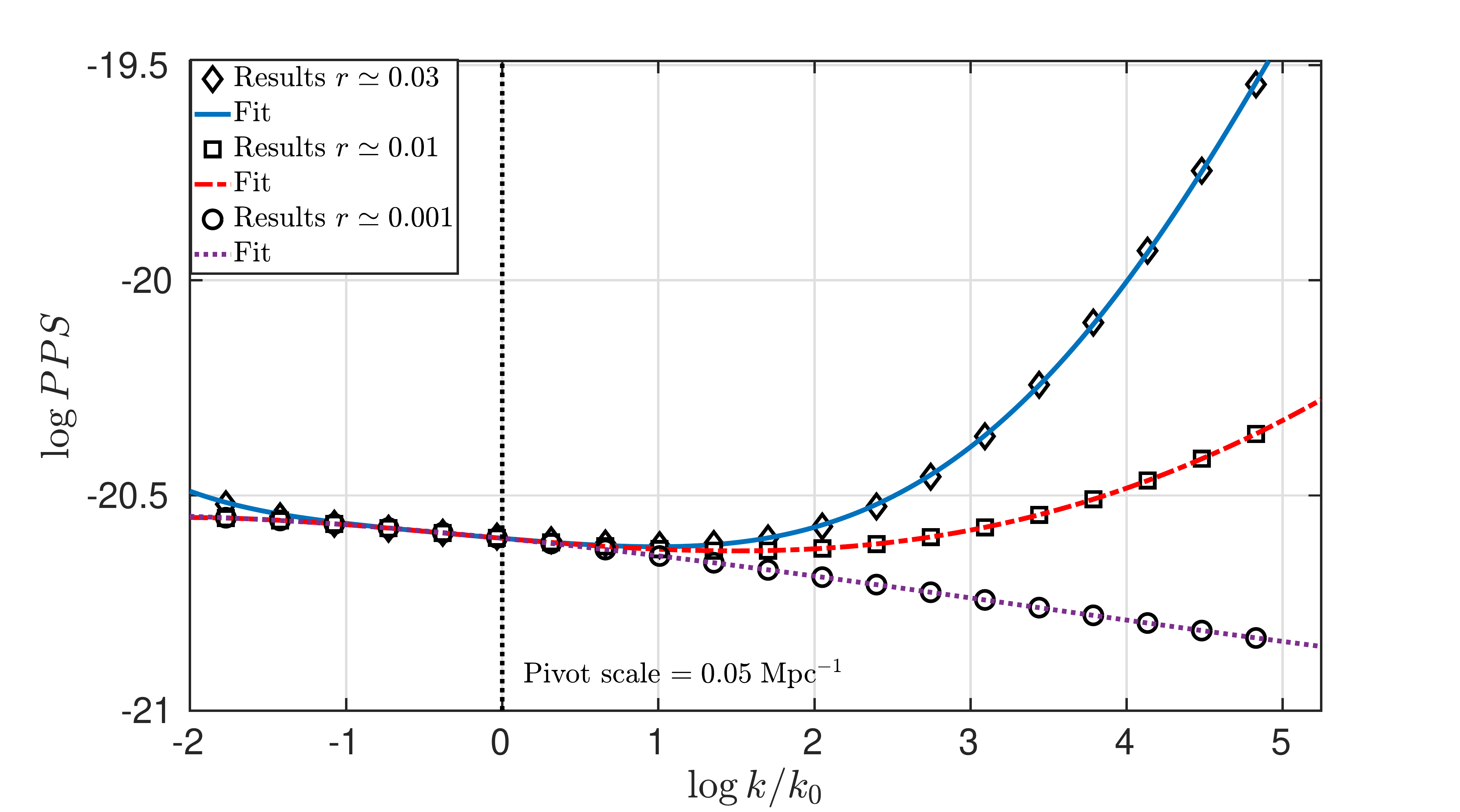}
\caption{Primordial power spectra of three  models. A model with a degree six polynomial potential that predicts $r\simeq 0.03$ (diamonds and blue line). A model with a degree six polynomial potential that predicts $r\simeq0.01$ (squares and red dash-dot)  and a model with a degree five polynomial potential that predicts $r=0.001$ (circles and purple dots). The pivot scale for all three is $k_0=0.05 \mathrm{Mpc^{-1}}$ and the results are overlayed at that scale for ease of comparison.\label{fig:PPS}}
\end{figure}

In Fig.~\ref{fig:PPS} the power spectra  generated by three inflationary models are shown. (1) A model with degree five polynomial potential that predicts $r\simeq 0.001$; (2) A model with degree six polynomial potential that predicts $r\simeq 0.01$; and finally (3) A model with degree six polynomial potential that predicts $r\simeq 0.03$.

Representing each coefficient $a_p$ as a function of the observables $(n_s,\alpha,\beta)$ and evaluating them at $(0.9687,0.008,0.02)$, leads to the following potential,
\begin{align}
	V=V_0\left[1-\sqrt{\frac{0.03}{8}}\phi -0.067\phi^2 +0.413\phi^3 -1.419\phi^4 +2.512\phi^5 -1.523\phi^6\right].
\end{align}
The values $n_s,\alpha,\beta$ that this model predicts deviate from the most likely values by $(0.06\%,10\%,19\%)$. However, the deviations are within the $95\%$ CL of all recent MCMC analyses.

\section{Conclusion}

We presented and discussed small field models of inflation with a degree six polynomial potential that predict $r\simeq 0.03$. The most likely of these models also predicts the most likely values of $n_s,\alpha,\beta$, within currently acceptable margins of error. The amount of coefficient tuning for these models was calculated. This article, along with its predecessors  \cite{Wolfson:2016vyx,Wolfson:2018lel}, demonstrates that an interesting range of values of  $r$, $0.001 \le r \le 0.03$ can be predicted by small field models of inflation that are consistent with the available CMB data.

\begin{acknowledgments}
The research of RB and IW was supported by the Israel Science Foundation grant no. 1294/16. IW would like to acknowledge Ido Ben-Dayan for useful discussions regarding the swampland conjecture and its cosmological implications.
\end{acknowledgments}

\end{document}